\documentclass[aps,prd,amsmath,amssymb,preprintnumbers]{revtex4}
\usepackage[a4paper,left=25mm,right=25mm,top=25mm,bottom=25mm]{geometry}
\usepackage{graphicx}
\usepackage{color}
\usepackage{subfigure}
\usepackage{fancyhdr}
\usepackage{multirow}
\usepackage{tikz}
\usepackage{dcolumn}
\usepackage{hyperref}
\usepackage{amsmath} 

\def\7{$\;$}
\def\l{\left}
\def\r{\right}
\def\be{\begin{equation}}
\def\ee{\end{equation}}
\def\bea{\begin{eqnarray}}
\def\eea{\end{eqnarray}}

\definecolor{rossoCP3}{cmyk}{0,0.88,0.77,0.40}

\begin{document}

\title{\Large \color{rossoCP3} Noether symmetry approach in the cosmological alpha-attractors}

\author{Narakorn Kaewkhao$^{1}$ and Thanyagamon Kanesom$^{2}$}
\affiliation{\vspace{3mm} Department of Physics, Faculty of Science, Prince of Songkla University, Hatyai 90112, Thailand\\
$^{1,2}$Email: {\rm naragorn.k@psu.ac.th, thanyagamon1995@gmail.com}}

\author{Phongpichit Channuie$^{3}$}
\affiliation{\vspace{3mm} School of Science, Walailak University, Thasala, Nakhon Si Thammarat, 80160, Thailand\\
$^{3}$Email: {\rm channuie@gmail.com}}

\begin{abstract}
In cosmological framework, Noether symmetry technique has revealed a useful tool in order to examine exact solutions. In this work, we first  introduce the Jordan-frame Lagrangian and apply the conformal transformation in order to obtain the Lagrangian equivalent to Einstein-frame form. We then analyse the dynamics of the field in the cosmological alpha-attractors using the Noether sysmetry approach by focusing on the single field scenario in the Einstein-frame form. We show that with a Noether symmetry the coresponding dynamical system can be completely integrated and the potential exhibited by the symmetry can be exactly obtained. With the proper choice of parameters, the behavior of the scale factor displays an exponential (de Sitter) behavior at the present epoch. Moreover, we discover that the Hubble parameters strongly depends on the initial values of parameters exhibited by the Noether symmetry. Interestingly, it can retardedly evolve and becomes a constant in the present epoch in all cases.
\\[2mm]
\end{abstract}

\maketitle \vskip 1pc

\section{Introduction} \label{ch1}

In modern cosmology, the mechanism of cosmic inflation seems conceivable. Inflation marks nowadays an inevitable ingredient when studying very early evolution of the universe. The reason stems from the fact that it solves most of the puzzles that plague the standard Big Bang theory, and simultaneously is consistent with recent observational data. In other words, it not only gives sensible explanations for the horizon, flatness, and relic abundant problems, but also provides us primordial density perturbation as seeds of the formation for a large-scale structure in the universe.  

Despite all its success, however, the underlying mechanism of the inflationary physics 
is still unknown. Recent observational data much flavors large field inflationary models with plateau-like inflaton
potentials. However, there are several different ways of constructing successful inflationary models. 
In the supergravity context, a model with plateau potentials was proposed by the authors of Refs.\cite{Goncharov:1984,Linde:2014hfa}. Here it describes a potential 
which is exponentially approaching a positive constant for
super-Planckian values of the inflaton field. Later on, a theory with a similar potential was realized 
as the Starobinsky model \cite{Starobinsky:1980te}, and then the Higgs inflation model with a similar potential was developed \cite{Salopek:1988qh,Bezrukov:2007ep}. It is worth noting that these models lead to nearly identical predictions, providing the best
fit to the latest Planck data \cite{Ade:2015lrj,Ade:2015xua}. Moreover, recent investigation shows the inflaton field can emerge as a composite state of a new strongly interacting gauge theory \cite{Channuie:2011rq,Bezrukov:2011mv}. More recently, a broad class of inflationary models, dubbed cosmological attractors \cite{Kallosh:2014rga,Kallosh:2013hoa,
Galante:2014ifa,Cecotti:2014ipa,Kallosh:2013daa,Yi:2016jqr,Eshaghi:2016kne}, yields very similar inflationary predictions.

Interestingly, the cosmological $\alpha$-attractors incorporate most of the existing inflationary models with plateau-like potentials including the Starobinsky model and some generalized versions of the Higgs inflation. Regarding the $\alpha$-attractors, the flatness of the inflaton potential is implemented and protected by the existence of a pole in the kinetic term of the scalar field. Moreover, at large-field values, any non-singular inflaton potential acquires a universal plateau-like form when performing the (conformal) transformation. Regarding the hyperbolic geometry and the flatness of the Kahler potential in the supergravity context, the universal behaviors of these theories make very similar cosmological predictions preserving good match to the latest cosmological observations \cite{Kallosh:2015zsa,Carrasco:2015uma,Carrasco:2015rva,Kallosh:2016ndd}. The successful extentions of these models can also describe dark energy/cosmological constant and supersymmetry breaking \cite{Ferrara:2014kva,Antoniadis:2014oya,Kallosh:2014via,Dall'Agata:2014oka,Kallosh:2014hxa,Lahanas:2015jwa,Kallosh:2015lwa,Carrasco:2015pla,Carrasco:2015iij}. 

The purpose of the present study is to analyse the dynamics of the field in the cosmological $\alpha$-attractors through the Noether sysmetry technique. Here we concentrate on the single field scenario. It is worth noting that this approach proved to be very useful not only to fix physically viable cosmological models with respect to the conserved quantities, e.g., couplings and potentials, but also to reduce dynamics and achieve exact solutions. Moreover, the existence of Noether sysmetries plays crucial roles when studying quantum cosmology \cite{Capozziello:2010}.

In the present paper, we consider the single-field model of cosmological $\alpha$-attractors. The structure of the paper is as follows: In Sec.\ref{ch2}, we introduce the model in the Jordan frame (JF) and transform the original action into the equivalent Einstein frame (EF) by applying the conformal transformation. In Sec.\ref{ch3}, we adopt the Noether symmetry approach to study the related dynamical systems obtained from a point-like Lagrangian. Here we can determine the form of the undefined potential of the action by imposing the Noether symmetry. We also compute the general solutions with the helps of the new coordinate system. Moreover, we compute the Hubble parameter and display plot as a function of time for some specific cases. Finally, we conclude our findings in the last section.

\section{The cosmological alpha-attractors in a nutshell}
\label{ch2}
As mentioned in Ref.\cite{Kallosh:2014rga}, the cosmological $\alpha$-attractors can be introduced in several inequivalent ways. To explain the structure of the cosmological attractors, we will start with a toy model with two-field scenario, $\chi$ and $\phi$. A precise form of the action for this model has been introduced in Ref.\cite{Kallosh:2014rga}. It has been shown that the theory is locally conformal invariant under the transformations and has a global $so(1,1)$ symmetry with respect to a boost between these two fields, preserving the value of $\chi^{2}-\phi^{2}$. One of these fields, say $\chi$, does not have any physical degrees of freedom associated to it and of course can be removed from the theory in several different ways. 

At the end of the day, the field $\chi$, called the conformal compensator or conformon, becomes the cutoff for possible values of another field $\phi$. In terms of the canonically normalized fields, the resulting theory is equivalent to a theory of gravity of a free massless canonically normalized field and a cosmological constant, see Ref.\cite{Kallosh:2014rga} for a detailed discussion. Here we first  introduce the Jordan-frame Lagrangian and apply the conformal transformation in order to obtain the Lagrangian equivalent to Einstein-frame form. Here we assume the cosmological scenario that our universe is described by the spatially-flat FLRW spacetime with metric signature $(+,-,-,-)$. Let us first consider
\bea \label{action}
\mathcal{S}_{\rm JF}=\int d^{4}x L_{\rm JF}=\int d^{4}x\sqrt{-g}\Bigg[-\Big(1-\frac{\phi^{2}}{6}\Big)\frac{R}{2} + \frac{\kappa(\alpha)}{2}g^{\mu\nu}\partial_{\mu}\phi\partial_{\nu}\phi-F(\phi)\Big(\frac{\phi^{2}}{6}-1\Big)^{2}\Bigg],   
\eea
where $\alpha$ is a free parameter and $\kappa(\alpha)$ is a function depending only on $\alpha$. However, in some specific cases it can be a constant or unity. Here we set the Planck mass $M_{P}=1$. With the field redefinition, $\phi = {\tilde \phi}/\sqrt{\alpha}$, the above Lagrangian density becomes
\bea \label{action2}
L_{\rm JF}=\sqrt{-g}\Bigg[-\frac{\Omega({\tilde \phi})}{2}R + \frac{\kappa(\alpha)}{2\alpha}g^{\mu\nu}\partial_{\mu}{\tilde \phi}\partial_{\nu}{\tilde \phi} -{\cal W}({\tilde \phi})\Bigg],   
\eea
where
\bea \label{action3}
\Omega({\tilde \phi})\equiv\Big(1-\frac{{\tilde \phi}^{2}}{6\alpha}\Big),\quad {\cal W}({\tilde \phi})\equiv F({\tilde \phi}/\sqrt{\alpha})\Big(\frac{{\tilde \phi}^{2}}{6\alpha}-1\Big)^{2}.   
\eea
Here and in what follows, we are going to transform th Jordan-frame Lagrangian to the canonical Einstein-frame form. In so doing, we next redefine the metric:
\bea \label{actio4}
g_{\mu\nu}\rightarrow \Omega({\tilde \phi})g_{\mu\nu} = \Big(1-\frac{{\tilde \phi}^{2}}{6\alpha}\Big)g_{\mu\nu},   
\eea
This brings the Jordan-frame Lagrangian to the Einstein-frame form
\bea \label{action25}
\mathcal{S}_{\rm EF}=\int d^{4}x L_{\rm EF}=\int d^{4}x\sqrt{-g}\Bigg[-\frac{R}{2} + \frac{1}{2}\left(\frac{\kappa(\alpha)}{\alpha \Omega} + \frac{3}{2}\left(\frac{\Omega'}{\Omega}\right)^{2} \right)g^{\mu\nu}\partial_{\mu}{\tilde \phi}\partial_{\nu}{\tilde \phi} - F({\tilde \phi}/\sqrt{\alpha})\Bigg],   
\eea
where the function of $\Omega$ is understood and a prime denotes differentiation with respect to the field ${\tilde \phi}$. Substituting $\Omega$ into Eq.(\ref{action25}), we discover
\bea \label{action12}
L_{\rm EF}=\sqrt{-g}\Bigg[-\frac{1}{2}R + \frac{\kappa(\alpha)}{2\alpha \Big(1-\frac{\phi^{2}}{6\alpha}\Big)^{2}}g^{\mu\nu}\partial_{\mu}\phi\partial_{\nu}\phi -F(\phi/\sqrt{\alpha})\Bigg],   
\eea
where we have droped tildes for convenience. However, at the end of the day, setting $\kappa(\alpha)=\alpha$ we exactly obtain the same result in the Einstein-frame form as proposed by several authors, see Refs.\cite{Kallosh:2014rga,Kallosh:2013hoa,Galante:2014ifa,Cecotti:2014ipa,Kallosh:2013daa,Yi:2016jqr,Eshaghi:2016kne} and references therein.

Here a parameter $\alpha$ determines the curvature and cutoff of the model. For inflation, the predictions depend on this parameter. At small cutoff, i.e. $\alpha \ll 1$, the resulting inflationary model featurs a plateau-like potentials with $n_{s}\approx 1-2/N$ and $r\approx 12\alpha/N^{2}$ with $N$ being a number of e-folding. For $\alpha =1$, the predictions coincide with those of the Starobinsky scenario and the Higgs one, i.e., $n_{s}=1-2/N$ and $r=12/N^{2}$. However, for large cutoff, i.e. $\alpha \gg 1$, the theory asymtotes to quadratic inflationary scenario, with $n_{s}\approx 1-2/N$ and $r\approx 8/N$. Interestingly, for intermediate values of $\alpha$, the predictions interpolate bewteen these two regimes covering the observed results of both Planck and BICEP2.

\section{Noether symmetry approach for alpha-attractors in the Einstein frame}
\label{ch3}
In this section we will in details consider the Noether symmetry approach. It is applied to determine the conserved quantities and constants of motion. After redefining the field, the action of the alpha-attractor models in the Einstein frame involving a real scalar field $\phi$ minimally coupled to gravity can be written as
\bea \label{EFac}
\mathcal{S}_{\rm EF}=\int d^{4}x L_{\rm EF}=\int d^{4}x\sqrt{-g}\Bigg[-\frac{R}{2} + \frac{\kappa(\alpha)}{2\alpha\Big(1-\frac{\phi^{2}}{6\alpha}\Big)^{2}}g^{\mu\nu}\partial_{\mu}\phi\partial_{\nu}\phi-F(\phi/\sqrt{\alpha})\Bigg].   
\eea
Having removed the coupling term, we come up with the non-canonical kinetic term for the scalar field $\phi$ which can be transformed into a canonical form by introducing the following field $\varphi(\phi)$ linked to $\phi$ via:
\bea \label{normfi}
\frac{1}{2}\tilde{g}^{\mu\nu}\partial_{\mu}\varphi(\phi)\partial_{\nu}\varphi(\phi) = \frac{1}{2}\left(\frac{\partial \varphi}{\partial \phi}\right)^{2}\tilde{g}^{\mu\nu}\partial_{\mu}\phi\partial_{\nu}\phi,   
\eea
where
\bea \label{normfi1}
\frac{\partial \varphi}{\partial \phi} \equiv \sqrt{\frac{\kappa(\alpha)}{\alpha\Big(1-\frac{\phi^{2}}{6\alpha}\Big)^{2}}}.   
\eea
However, we can easily obtain the canonically normalized variable $\varphi$  by solving the equation:
\bea \label{norm}
d\varphi = \int \frac{\beta(\alpha)}{\Big(1-\frac{\phi^{2}}{6\alpha}\Big)} d\phi,   
\eea
so that we simply find $\phi =\sqrt{6\alpha}\beta(\alpha)\tanh(\varphi/\sqrt{6\alpha})$ and $\beta(\alpha)=\sqrt{\kappa/\alpha}$. In terms of this new variable, it is calculable and rather straightforward to proceed further. The resulting action written in terms of the canonically normalized field is given by
\bea \label{EFnor}
 L_{\rm EF}=\int d^{4}x\sqrt{-g}\Bigg[-\frac{R}{2} + \frac{1}{2}g^{\mu\nu}\partial_{\mu}\varphi\partial_{\nu}\varphi -  F(\varphi)\Bigg].
\eea
The Euler-Lagrange equation for a normalized scalar field $\varphi$ is given by
\bea \label{EL}
\ddot{\varphi} + 3H\dot{\varphi} + F'(\varphi) = 0,
\eea
where a prime denotes derivatives with respect to $\varphi$ and $F(\varphi)$ can be viewed as a potential of the model. Our purpose here is to analyse the cosmological aspects of the model in the next subsection through the Noether symmetry approach. In so doing, we restrict ourself to a flat FLRW cosmology with scale factor $a(t)$ and in this case after performing an integration by parts the point-like Lagrangian of Eq(\ref{EFnor}) reads
\bea\label{Lapo1}
L_{\rm EF}= -3a\dot{a}^{2} + \frac{a^3\dot{\varphi}^{2}}{2} - a^{3}F(\varphi) ,
\eea
where the function of the Lagrangian is understood. The configuration space for such a Lagrangian is ${\cal Q}=(a,\,\varphi)$ and then cosmological dynamics can be implemented on such a two-dimensional minisuperspace. Note that the extension of the above action may be possible for a non-linear $f(R)$ theory of gravity. Another word of saying is that the attractors could happen in the non-linear gravity, see for example \cite{Yi:2016jqr}.  

\subsection{Noether symmetry approach}
\label{ssch3}

Regarding the preceeding consideration, we obtain the point-like Lagrangian Eq.(\ref{Lapo1}) written in terms of the scale factor $a$ and a scalar field $\varphi$. Notice that they play the role of independent dynamical variables. Our manipulation here is that the Lagrangian is constructed in such a way that its variation with respect to $a$ and $\varphi$ yields the correct equations of motion. The form of the potential $F(\varphi)$ apprearing in the action Eq.(\ref{Lapo1}) can be in priciple determined by demanding that the Lagrangin admits the desired Noether symmetry \cite{Demianski:1992tu,Capozziello:1993vr,Capozziello:1996ay,Capozziello:2007iu,Capozziello:2008ch}. As is well known, the Noether symmetry approach is a powerful tool to obtain the solution for a given Lagrangian. Besed on the Noether theorem, if there exists a vector field $X$, for which the Lie derivative of a given Lagrangian $L$ vanishes, i.e. ${\cal L}_{X}L=XL=0$, the Lagrangian admits a Noether symmetry and thus yields a conserved current. The configuration space of the Lagrangian is ${\cal Q}(a,\,\varphi)$ and the corresponding tangent space is $T{\cal Q}=(a,\,\varphi,\,\dot{a},\,\dot{\varphi})$. Therefore, the generic infinitesimal generator of the Noether symmetry is
\bea
X= \delta \frac{\partial}{\partial a}+\lambda \frac{\partial}{\partial \varphi} + \dot{\delta}\frac{\partial}{\partial \dot{a}}+\dot{\lambda}\frac{\partial}{\partial \dot{\varphi}},\label{xop1}
\eea
where dots represent derivatives with respect to time, and $\delta,\,\lambda$ depend on $a$ and $\phi $. As of the previous section, it is straighforward to figure out the constant of motion corresponding to such symmetry. Indeed, $L_{X}{\cal L}=0$ can be explicitely rewritten as 
\bea \label{xop}
{\cal L}_{X} L= \l(\delta \frac{\partial L}{\partial a}+ \dot{\delta}\frac{\partial L}{\partial \dot{a}}\r)+\l(\lambda \frac{\partial L}{\partial \varphi} +\dot{\lambda}\frac{\partial L}{\partial \dot{\varphi}}\r) =0,
\eea
where $\delta$ and $\lambda$ are functions only of $a$ and $\phi$. By applying the symmetry condition Eq.(\ref{xop}) to Eq.(\ref{Lapo1}), with respect to the vector field (\ref{xop1}), we obtain a system of coupled partial diffential equations as follows: 
\bea
\delta+2a\frac{\partial\delta}{\partial a}&=&0,\label{e1}\\
6\frac{\partial\delta}{\partial \varphi} - a^{2}\frac{\partial\lambda}{\partial a}&=&0,\label{e2}\\
\frac{3\delta}{2} +a\frac{\partial \lambda}{\partial \varphi}&=&0,\label{e3}\\
3\delta F + \lambda aF'&=&0.\label{e4}
\eea
In the following we are going to solve the system of coupled partial differential equations given above. Following Ref.\cite{Demianski:1992tu,Capozziello:1993vr,Capozziello:1996ay,Capozziello:2007iu,Capozziello:2008ch,Darabi:2015gaa,Sk:2016dkj,deSouza:2013uu,Paliathanasis:2014rja}, we consider $\delta$ and $\lambda$ as separable functions of $a$ and $\varphi$, i.e., $\delta\equiv A_{1}(a)B_{1}(\varphi)$ and $\lambda\equiv A_{2}(a)B_{2}(\varphi)$. Hence, we find for Eq.(\ref{e1})
\bea
\delta=\delta(a,\phi) = \frac{1}{a^{1/2}}B_{1}(\varphi), \label{alpha}
\eea
where $B_{1}(\varphi)$ depends only on the field $\varphi$. Equation (\ref{e4}) gives $\delta =  -a\lambda\psi$ with $\psi \equiv F'/3F$. Plugging these relations into Eq.(\ref{e1}), we obtain 
\bea
\lambda=\lambda(a,\phi) = \frac{1}{a^{3/2}}B_{2}(\varphi), \label{beta}
\eea
where $B_{2}(\varphi)$ depends only on the field $\varphi$. From Eq.(\ref{e2}) together with Eq.(\ref{e3}), we find a solution for $B_{1}(\varphi)$ given by
\bea
B_{1}(\varphi) =- A\sinh \Big(\sqrt{\frac{3}{8}}\varphi\Big), \label{betaso}
\eea
where $A$ is an integration constant. Therefore, the general solution for $\delta(a,\phi)$ reads
\bea
\delta=\delta(a,\varphi) =- \frac{A}{a^{1/2}}\sinh \Big(\sqrt{\frac{3}{8}}\varphi\Big). \label{betafull}
\eea
Plugging the solution for $\delta$ into Eq.(\ref{e3}), we obtain the solution for $\lambda$ given by
\bea
\lambda= \lambda(a,\phi) = \frac{3}{2}\frac{A}{a^{3/2}}\cosh \Big(\sqrt{\frac{3}{8}}\varphi\Big) .
\eea
Finally, substituting $\delta$ and $\lambda$  into Eq.(\ref{e4}), the Noether symmetry selects the potential $F$ of the form
\bea
F(\varphi) = F_{0}\cosh^{2}\Big(\sqrt{\frac{3}{8}}\varphi\Big) .
\eea
with $F_{0}$ being a constant. In the following subsection, we determine the general solutions of the system and compute some parameters.
\begin{figure*}
	\begin{center}
		\includegraphics[width=0.45\linewidth]{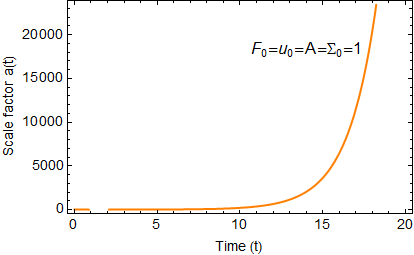}
		\includegraphics[width=0.5\linewidth]{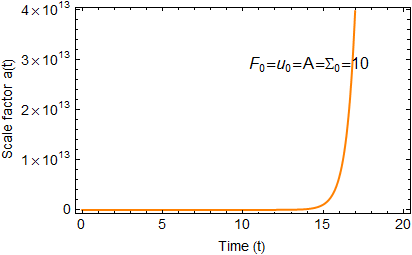}
		\includegraphics[width=0.45\linewidth]{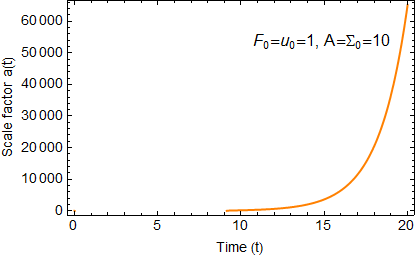}
		\includegraphics[width=0.5\linewidth]{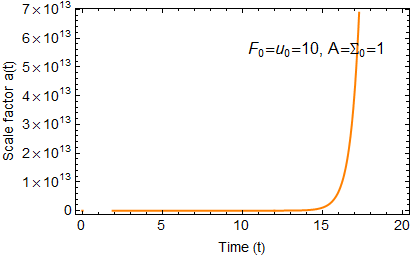}
		\caption{The behavior of the scale factor $a(t)$  for varius conditions: the upper-left panel $ F_{0}= u_{0} = A =\Sigma_{0}= 1$, the upper-right panel $ F_{0}= u_{0} = A =\Sigma_{0}= 10$, the lower-left panel $ F_{0}= u_{0}=1, A =\Sigma_{0}= 10$ and the lower-right one $ F_{0}= u_{0}=10, A =\Sigma_{0}= 1$.}
		\label{fig1}
	\end{center}
\end{figure*}

\subsection{The solutions in general case}
\label{ssch2}
For any given Lagrangian, the Noether symmetry approach is a useful
tool in finding the solutions including the present one. In this approach, the technique is concerned with finding
the cyclic variables related to the conserved quantities and consequently
reducing the dynamics of the system to a analytically calculable manner. In this subsection, we shall pursue these new variables to examine the potential such that the corresponding Lagrangian exhibits the desired symmetry.

\begin{figure*}
	\begin{center}
		\includegraphics[width=0.6\linewidth]{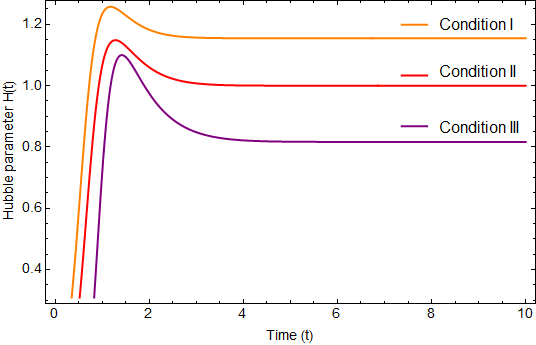}
		\caption{The Hubble parameter $H(t)$ as a function of time for the upper line (Condition I) $F_{0}= 4, u_{0} = 1, A = 1, \Sigma_{0}= 1$, the middle one (Condition II) $F_{0}= 3, u_{0} = 1, A = 1, \Sigma_{0}= 1$ and the lower one (Condition III) $F_{0}= 2, u_{0} = 1, A = 1, \Sigma_{0}= 1$.}
		\label{fig2}
	\end{center}
\end{figure*}

Here we start by rewriting the point-like Lagrangian (\ref{EFnor}) in other system of coordinates, namely $(u,w)$. We deduce that there must exist a coordinate transformation in the configuration space in which one of these variables is cyclic. Such a transformation obeys the following system of differential equations:
\bea
\delta \frac{\partial u}{\partial a}+ \lambda\frac{\partial u}{\partial \varphi}&=&0,\label{c1}\\
\delta \frac{\partial w}{\partial a}+ \lambda\frac{\partial w}{\partial \varphi}&=&1.\label{c2}
\eea
Note that $u$ and $w$ are the new coordinates connected to the old ones, i.e., $a$ and $\varphi$. In this system, $w$ is the cyclic coordinate. We find a particular solution of this system:
\bea
u &=& a^{3}\cosh\Big(\sqrt{\frac{3}{8}}\varphi\Big)^{2},\label{cy1}\\
w &=& \frac{2}{3A}a^{3/2}\sinh\Big(\sqrt{\frac{3}{8}}\varphi\Big),\label{cy2}
\eea
or equivalently
\bea
a &=& \Big(u - (9/4)A^{2}w^{2}\Big)^{1/3},\label{cye1}\\
\varphi &=& 2\sqrt{2/3}\,\,{\rm arcsinh}\Big(\frac{3Aw/2}{\sqrt{u - (9/4)A^{2}w^{2}}}\Big).\label{cye2}
\eea
The point-like Lagrangian rewritten in terms of these new corrdinates takes the form
\bea\label{L1}
\mathcal{L}_{\rm EF}= -\frac{1}{3}\frac{\dot{u}^{2}}{u} + 3 A^{2}\dot{w}^{2} - u{\cal F}_{0},
\eea
where $w$ is the cyclic variable. The Euler-Lagrange equations for Eq.(\ref{L1}) read
\bea \label{solse}
6A^{2}\ddot{w}(t) &=&0,\label{sow1e}\\   
2u(t)\ddot{u}(t) - \dot{u}(t)^{2} - 3F_{0}u(t)^{2} &=&0 .\label{sow2e}
\eea
The Eq.(\ref{sow1e}) can be trivially integrated to obtain $w(t) = \Sigma_{0} t+C$ where a factor $6A^{2}$ is obsorbed in $\Sigma_{0}$ and we can set $C = 0$ without loosing of generality. It turns out that the general solution of the second differential equation is particularly given by 
\bea\label{solu}
u(t) = u_{0} \cosh\Big(\frac{1}{2}\sqrt{3F_{0}}t\Big)^2,
\eea
where $u_{0}$ is an integration constant. The substitution of the functions $a = a(u, w)$ and $\varphi = \varphi(u, w)$ into Eqs.(\ref{cye1}) and (\ref{cye2}) yields the scale factor and the scalar field which evolve as
\bea
a(t) &=& \Big(u_{0} \cosh\Big(\frac{1}{2}\sqrt{3F_{0}}t\Big)^2 - (9/4)A^{2}\Sigma_{0}^{2} t^{2}\Big)^{1/3},\label{cye1e}\\
\varphi(t) &=& 2\sqrt{2/3}\,\,{\rm arcsinh}\Big(\frac{3A\Sigma_{0} t/2}{\sqrt{u_{0} \cosh\Big(\frac{1}{2}\sqrt{3F_{0}}t\Big)^2 - (9/4)A^{2}\Sigma_{0}^{2} t^{2}}}\Big).\label{cye2e}
\eea
The scale factor behavior is shown as Fig.(\ref{fig1}). We find that its bahaviors are very sensitive to the values of $F_{0}$ and $u_{0}$.  It displays a power-law domination for small values of $F_{0}$ and $u_{0}$ illustating by the left panels of  Fig.(\ref{fig1}). Here we use $ F_{0}= u_{0} = A =\Sigma_{0}= 1$ and $F_{0}= u_{0} < A =\Sigma_{0}= 10$. Interestingly, for large values of $F_{0}$ and $u_{0}$, i.e. $F_{0}=u_{0}\gg 1$,  the scale factor has an exponential behavior illustating by the right panels of  Fig.(\ref{fig1}). Using Eq.(\ref{cye2e}) allows us to transform the field $\varphi$ to the original one, $\phi$:
\bea
\phi(t) =\sqrt{6\alpha}\beta(\alpha)\tanh\left[\frac{2}{3\sqrt{\alpha}}\,\,{\rm arcsinh}\Big(\frac{3A\Sigma_{0} t/2}{\sqrt{u_{0} \cosh\Big(\frac{1}{2}\sqrt{3F_{0}}t\Big)^2 - (9/4) A^{2}\Sigma_{0}^{2} t^{2}}}\Big)\right].\label{cp}
\eea
It is very trivial in this case to determine the Hubble parameter to obtain
\bea
H(t) =\frac{2 \left(\sqrt{3} \sqrt{F_0} u_0 \sinh \left(\sqrt{3} \sqrt{F_0} t\right)-9 A^2 \Sigma_{0}^2 t\right)}{-27 A^2 \Sigma_{0}^2 t^2+6  u_0 \cosh \left(\sqrt{3} \sqrt{F_0} t\right)+6  u_0}.
\eea
Its behavior is shown in Fig.(\ref{fig2}). Notice that the Hubble parameters strongly depends on the initail values of $F_{0},u_{0},A$ and $\Sigma_{0}$ exhibited by the Noether symmetry. Precisely, it is constant in the present epoch in all cases. Moreover, by properly choosing the initial values of the parameters, it can evolve last long before its values become a constant at the present epoch

\section{Conclusions}
In this work, we introduce the Jordan-frame Lagrangian and apply the conformal transformation in order to obtain the Lagrangian equivalent to Einstein-frame form. We then analyse the dynamics of the field in the cosmological alpha-attractors using the Noether sysmetry technique. Here we concentrate on the single field scenario in the Einstein-frame form. We show that by using a Noether symmetry the coresponding dynamical system can be completely integrated and the potential exhibited by the symmetry can be exactly obtained.

The scale factor behavior is very sensitive to the initail values of $F_{0}$ and $u_{0}$. We discover that it displays a power-law domination for small values of $F_{0}$ and $u_{0}$. However, it shows an exponential (de Sitter) behavior for large values of $F_{0}$ and $u_{0}$ at the present time. The Hubble parameter can be easily specified in this work. Its behavior is shown in Fig.(\ref{fig2}). Notice that the Hubble parameters strongly depends on the initail values of $F_{0},u_{0},A$ and $\Sigma_{0}$ exhibited by the Noether symmetry. Precisely, it is constant in the present epoch in all cases. Moreover, by properly choosing the initial values of the parameters, it can retardedly evolve before its values become a constant at the present epoch. Last but not the least, the extension of the present analysis may be possible for a non-linear $f(R)$ theory of gravity. Likewise, the quantum vertion of the present analysis can be possibly achieved by following Ref.\cite{DeWitt,Hartle} based upon on the Wheeler-DeWitt (WDW) equation.

\end{document}